# Critical influence of target-to-substrate distance on conductive properties of LaGaO$_3$/SrTiO$_3$ interfaces deposited at 10$^{-1}$ mbar oxygen pressure

C. Aruta[1], S. Amoruso, G. Ausanio, R. Bruzzese, E. Di Gennaro, M. Lanzano,

F. Miletto Granozio, Muhammad Riaz, A. Sambri, U. Scotti di Uccio, and X. Wang

*CNR-SPIN and Dipartimento di Scienze Fisiche, Complesso Universitario di Monte Sant'Angelo, Via Cintia, I-80125 Napoli, Italy*

**Abstract**

We investigate pulsed laser deposition of LaGaO$_3$/SrTiO$_3$ at 10$^{-1}$ mbar oxygen background pressure, demonstrating the critical effect of the target-to-substrate distance, $d_{TS}$, on the interface sheet resistance, $R_s$. The interface turns from insulating to metallic by progressively decreasing $d_{TS}$. The analysis of the LaGaO$_3$ plume evidences the important role of the plume propagation dynamics on the interface properties. These results demonstrate the growth of conducting interfaces at an oxygen pressure of 10$^{-1}$ mbar, an experimental condition where a well-oxygenated heterostructures with a reduced content of oxygen defects is expected.

---

[1] Electronic mail: carmela.aruta@spin.cnr.it

The discovery of a two-dimensional electron gas (2DEG) at the interface between LaAlO$_3$ (LAO) and SrTiO$_3$ (STO)[1] raised the question whether similar properties can be found in samples where LAO is substituted by a different overlayer. Recently, interfaces between Lanthanum Gallate (LGO) and STO were also shown to host a 2DEG.[2] LGO and LAO compounds share most physical properties, in terms of both structure and electronic properties. To date, both experimental data and first-principles calculations seem to indicate that LAO/STO and LGO/STO interfaces also share similar microstructure and electronic properties,[2,3] such as the polar discontinuity at the interface, which makes appropriate the description of the 2DEG formation in terms of an electronic reconstruction.[4] Nevertheless, mechanisms involving oxygen vacancies already considered as possible sources of the LAO/STO interface conductivity,[5,6] may also be envisaged for LGO/STO. In fact, each oxygen vacancy acts in STO as a donor, bringing two electrons in the conduction band. When the interfaces are fabricated at low oxygen pressure, oxygen vacancies can be either directly formed in the STO substrate, prior to deposition, or induced by the interaction with the growing film.[7,8]

Up to now, layer-by-layer growth of conductive interfaces has been only carried out at oxygen pressure of $10^{-6}$ - $10^{-2}$ mbar.[5,6,9,10] Post-deposition treatments, e.g. in 0.2 bar of O$_2$ at ≈530 °C, have been proposed as a viable route to decrease the amount of oxygen vacancies.[10] However, such a post-growth process imposes some constrains in view of deposition of more complex heterostructures and multilayers. It is also worth noticing that the theoretically predicted electronic phase separation has been recently observed in LAO/STO only when the interfaces are grown at a high oxygen pressure ($10^{-2}$ mbar),[9] a regime not previously explored because too close to the three-dimensional growth mode. Opening the route to an even higher pressure regime, where the interfaces are conducting and the growth is two-dimensional (2D), might be very useful to explore new interfacial electronic and magnetic phenomena.

However, previous attempts to fabricate polar/non polar interfaces at an even higher pressure of $10^{-1}$ mbar resulted in insulating samples,[12] even though this regime is fully compatible with homo- and hetero-epitaxy of perovskites (in particular, of LAO and LGO).[11,12] Here we demonstrate that sample properties critically depend on the target-to-substrate distance. As a result, we show that conducting LGO/STO interfaces can actually be grown at an oxygen pressure of $10^{-1}$ mbar. The LGO ablation plume propagation into the background gas is analyzed to evidence the direct influence of the ablation plume features on the interface properties.

LGO films were grown by PLD on a $TiO_2$-terminated (001) STO substrate held at 800 °C in a $10^{-1}$ mbar oxygen atmosphere. A KrF excimer laser beam (248 nm, 25 ns duration full width half maximum) was focused on a stoichiometric target. The laser spot-size and fluence were $1.5 \times 10^{-2}$ cm$^2$ and 2.5 J/cm$^2$, respectively. The growth process was monitored by high pressure reflection high energy electron diffraction (RHEED). The ablation plume dynamics was investigated by fast photography.[13,14] Plume analyses at $10^{-3}$ and 1 mbar were also performed to elucidate the effect of background gas pressure.

Fig. 1(a) reports the temperature-dependent sheet resistance, $R_s(T)$, of 12 unit cells (uc) thick LGO/STO heterostructures grown at an oxygen pressure of $\approx 10^{-1}$ mbar, for three different target-to-substrate distances, $d_{TS}$. The interface grown at $d_{TS}$=35 mm is insulating, as previously reported,[12] but conductivity is progressively enhanced as $d_{TS}$ decreases. $R_s(T)$ observed for $d_{TS}$=30 mm is similar to that obtained in samples deposited in an oxygen pressure range of $10^{-4}$-$10^{-2}$ mbar.[2] Further reduction of $d_{TS}$ was hindered by the substrate heater shading the incoming laser beam. Nevertheless, the experimental findings of Fig. 1 illustrate the critical dependence of $R_s$ on $d_{TS}$, which is eventually related to variation in the plume characteristics, as discussed in this letter.

As an example of the growth process at $10^{-1}$ mbar, the oscillations of the RHEED diffracted pattern intensity registered during the growth of a LGO film are reported in the left panel of Fig. 1(b). Similar RHEED oscillations were also observed at lower oxygen pressure.[2] On the base of the clear 2D RHEED pattern at the end of deposition (see Fig.1(b), upper panel), we can rule out significant variation of the surface ordering during the growth process. Atomic Force Microscopy (AFM) analysis of the samples shows a smooth step-terrace structure, demonstrating that the LGO surface is characterized by high flatness. As an example, Fig. 1(c) reports an AFM image of the sample deposited at $d_{TS}$=30 mm, in which terraces separated by 1 uc steps are clearly discerned (see upper panel). In concert with the RHEED oscillations, this indicates that the 2D growth mode is dominant. Our experimental findings, therefore, suggest that at a pressure of $\approx 10^{-1}$ mbar: (i) deposition conditions exist where the growth proceeds in the layer-by-layer mode; (ii) variation of the target-substrate distance, and hence of the characteristics of the plume species impinging on the STO substrate, may significantly affect the sample interfacial conductivity. The critical influence of $d_{TS}$ on the LGO growth process and interface conductivity is demonstrated by the dramatic variation of the LGO/STO sheet resistance at room temperature, $R_s^*$ and of the deposition rate, $\delta$, obtained by RHEED monitoring as summarized in Table I.

The possibility to grow conductive STO-based oxide heterostructures at $10^{-1}$ mbar is particularly attractive in view of obtaining samples where intrinsic mechanisms of the interfacial conduction might dominate. In fact, at this pressure level: i) La-subplantation due to energetic plume species impacting on the STO substrate is negligible, as a consequence of the drastic reduction of the kinetic energy of the ablated species at the deposition distance;[12,15] ii) larger oxidation the LGO overlayer limits diffusion of oxygen from the STO substrate to the growing film.[9,12,15]

The following analysis of plume expansion dynamics allows elucidating the effect of background pressure and target-to-substrate distance. Fig. 2 reports single-shot, fast images of the LGO plume emission collected by intensified-charge-coupled device (ICCD) at three different pressure levels, for two succeeding temporal delays, $t$, after the laser pulse. In Fig. 2, the image intensity is multiplied by an appropriate factor, reported in parenthesis in the bottom-left corner of each panel, in order to compensate the progressive reduction of the intensity with time, as a consequence of the plume expansion, and to facilitate the comparison. In these experiments, the heated substrate was located at an intermediate distance of 33 mm.

The images of Fig. 2 show a significant influence of oxygen pressure on plume propagation dynamics that eventually affects growth dynamics and film characteristics. The images at $10^{-3}$ mbar are representative of the low-pressure deposition conditions usually employed in PLD interface deposition. One can observe that the plume expands freely, and its front reaches the substrate already for $t \approx 4$ μs. The fast plume expansion is accompanied by a significant reduction of the plume intensity with $t$, due to a progressive decrease of plume density and temperature. The images at $10^{-1}$ mbar allow evidencing the important variation induced on plume dynamics in the new regime explored here. At $10^{-1}$ mbar, the plume-background gas interaction causes a significant plume deceleration, and a much intense plume emission, due to the formation of a large number of excited species as a consequence of the plume interaction with the background gas.[12,15] These excited species are mainly located at the plume front and directly impact on the STO substrate, thus influencing the film growth. Finally, at the still larger pressure of ≈1 mbar, the plume front is strongly braked and eventually stopped before reaching the substrate.[16,17] In this last case, the plume species can reach the substrate only through diffusion into the background gas.[18]

The above discussed experimental observations show a crucial dependence of the plume properties on oxygen pressure and on distance from the target surface, which in turn influences the growth process and the conduction properties of the LGO/STO interfaces (see Fig. 1 and Table I). The plasma plume propagation in the ambient gas is usually analyzed by using position-time, *R-t,* plots of the leading edge of the plume emission along the normal to the target surface.[13-16] The plume front position *R* is defined as the distance at which the integral of the emission intensity attains a value equal to 95% of the total emission along the direction, *z*, normal to the target surface.[14] The complex distance-related-pressure influence of the ambient gas on the propagation dynamics and energetic state of the expanding plume has been illustrated by several theoretical and experimental analyses,[13-17] and the use of dimensionless variables has been shown to be advantageous in order to rationalize and interpret its behavior.[16,17] To identify the plume conditions resulting in the different behavior of the sheet resistance of Fig. 1(a), in the following we resort to the Predtechensky and Mayorov (PM) model of plume propagation,[19] which allows taking into account the variable background gas density with distance from the target induced by substrate heating at high temperature.[14,20]

The PM model is based on the balance between the plume linear momentum variation and the external pressure force. The model considers plume and adjoint background gas as a hemispherical thin layer of radius *R* moving at velocity *u* and experiencing the force due to the background gas pressure *p*. Then, the equations of motion for *R* and *u* read as follows:

$$\frac{d}{dt}\{[M_p + M_g(R)]u\} = -2\pi R^2 p\; ; \qquad u = \frac{dR}{dt} \qquad (1)$$

where $M_p$ is the confined plume mass. $M_g(R) = \int_0^R 2\pi r^2 \rho(r)dr$ is the mass of the background gas swept away by the expanding plume at a distance *R* and time *t*, where *ρ(r)* is the background density profile as a function of the radial coordinate $r$. The initial conditions are $R(t=0)=0$ and $u(t=0)=u_0$. As a consequence of substrate heating, the plume front encounters a gas with a variable density during its motion along the normal to the target surface. This is taken into account by using a background density profile of the form $f(r)=\rho(r)/\rho_0=(1+\beta r)^{-1}$, where $\beta$ is a constant which depends on substrate temperature and target-to-substrate distance, and $\rho_0$ is the density at $r=0$.[20] The plume dynamics can be analyzed in terms of the following dimensionless variables: time, $\tau=c_0 t/a_0$; position, $\xi=R/a_0$; velocity $\eta=u/c_0$. The parameter $a_0=(3M_p/2\pi\rho_0)^{1/3}$ is a characteristic distance which depends on the experimental conditions, while $c_0=(p/\rho_0)^{0.5}$ is a characteristic velocity equal to 278.2 m/s for oxygen background gas.

The observed plume dynamics in dimensionless form is shown in Fig. 3(a). In dimensionless coordinates Eq.(1) reads:

$$\frac{d}{d\tau}\left\{\left[1+3\int_0^{\xi(\tau)} f(\zeta)\zeta^2 d\zeta\right]\eta\right\} = -3\xi^2 \qquad (2)$$

Eq.(2) has been used to fit the experimental data of Fig. 3(a) with the initial conditions $\xi(\tau=0)=0$ and $\eta(\tau=0)=\eta_0=37.75$, which corresponds to a free-plume front velocity $u_0=1.05\times10^4$ m/s. The model predictions describe fairly well the experimental data, and allow identifying the various stages of the plume dynamics in the different experimental conditions. Initially, a free-expansion ($\xi=\eta_0 \tau$, dash-dot line in Fig. 3(a)) occurs until $\xi_{fp}\approx1.26$, which corresponds to the physical condition of an adjoint background gas mass $M_g$ equal to the plume mass $M_p$. Thereafter, the plume expansion begins to slow down, and at a certain stage it follows a shock-wave (SW) behavior ($\xi\propto\tau^{2/5}$), as shown in Fig 3(b). At later stages, the plume front continues to decelerate, and eventually stops at $\xi_{st} \approx 4.3$. From the stopping distance, $R_{st} \approx \xi_{st} a_0 \approx 22$ mm observed at $p=10^0$ mbar, we estimate $M_p \approx 5\times10^{-10}$ kg.

As $\xi \propto R\, p^{1/3}$, the various plume propagation regimes observed in Fig. 3(a) are associated to different choices of target-to-substrate distance, $d_{TS}$, and gas pressure $p$, clearly evidencing the distance-related pressure dependence of the plume propagation. In this respect, it is worth noticing that for deposition on a heated substrate, any variation of $d_{TS}$ also influences the background density profile. In our experimental conditions (800 °C, 30 mm $< d_{TS} <$ 35 mm), the ratio $\rho(d_{TS})/\rho_0$ is $\approx 27\text{-}28\,\%$ at $d_{TS} = 30\text{-}35$ mm, while the adjoint background mass $M_g(d_{TS})$ variation is only $\approx 15\%$ by passing from $d_{TS}=30$ mm to $d_{TS}=35$ mm. Therefore, the plume propagation dynamics reported in Figs. 2 and 3(a), obtained for $d_{TS}=33$ mm, can be reliably exploited to gain information on the plume propagation regimes associated to the conditions used for the fabrication of the LGO/STO heterostructures previously discussed (see Fig. 1 and Table I, e.g.). The dimensionless coordinates $(\tau,\xi)$ corresponding to the fabrication conditions, i.e. the target-to-substrate distance and the arrival time of the plume at the substrate position, are shown as symbols in Fig. 3(b). One can observe that for the three shorter distances used, the growth process takes place in the SW-like regime, while at $d_{TS}=38$ mm the plume has already turned to a slightly more slowed propagation regime. Moreover, as $d_{TS}$ increases a gradual reduction of the deposition rate, $\delta$, and maximum plume front velocity at the substrate position, $u_s$, occur as a consequence of the progressively larger braking effect of the background gas, as reported in Table I ($u_s = \eta_s c_0$, $\eta_s$ being the dimensionless velocity at the substrate distance estimated by the fit to PM model in Fig. 3(a)). We observe that the maximum kinetic energy of the plume cations impacting the substrate, $KE_s = \frac{1}{2} m\, u_s^2$ (where $m$ is Ga or La mass), changes from $\approx 1\text{-}2$ eV at $d_{TS}=30$ mm to $\approx 0.6\text{-}0.9$ eV at $d_{TS}= 35$ mm. As a final remark, we observe that the dragging action of the expanding plume driving the background oxygen molecules towards the substrate occurs for $\xi > \xi_{fp}$. This condition corresponds to a distance of ≈15 mm from the target surface at $p$=10$^{-1}$ mbar, but it increases to ≈70 mm for $p$=10$^{-3}$ mbar.

The previous analysis suggests that appropriate tuning of the parameters allows selecting conditions where the SW-like regime facilitate the deposition of well-oxygenated, 2D interfaces. The SW-like regime is an experimental condition which determines a higher internal energy of the plume during the growth process,[16,17] while favoring ablated species oxidation.[12,15,21] In particular, a larger content of excited and oxidized plume species reaches the substrate in such a condition. The upper electronic levels of these excited species are at 2-3 eV above the ground state.[12] This internal excitation energy is eventually supplied to the growing film during deposition.[22] This energy is comparable with the maximum surface diffusion barrier energy at high coverage reported in ref. 23 and, when released to the growing film, can promote surface diffusion, thus resulting in the observed 2D growth (see Fig. 1(b)). Nonetheless, the achievement of conductive interfaces critically depend on the target-to-substrate distance, as shown in Fig. 1(a). In particular, the variation of $KE_s$ with $d_{TS}$ points to an important role of the particles kinetic energy on the final interface conductivity, an aspect which has not been fully considered so far. A reduction of the cation maximum kinetic energy below ≈1eV results in a less conductive or insulating interface (see Fig. 1(a) and Table I).

In conclusion, we used a simultaneous analysis of the film growth and of the laser ablated plume dynamics to demonstrate that suitable experimental conditions exist for the deposition of conducting LGO/STO interfaces at an oxygen pressure of ≈10$^{-1}$ mbar, a situation where an optimal oxidation of the film is expected. This is attained in a SW-like regime of the plume propagation, which results in 2D growth even at such pressure, and locating the substrate at a position where the maximum kinetic energy of the impinging species is still of the order of ≈1 eV. In this situation the interfacial conducting properties should be free from extensive oxygen defects contribution. The possibility of releasing the constraints of low oxygen pressure growth can allow deeper understanding of the role of the various mechanisms contributing to interface conductivity. This opens new perspectives in the comprehension of the subtle mechanisms underlying the formation of the electron gas developing at polar/non-polar interfaces, and, more generally, of the growth process of well oxygenated and ordered oxide interfaces, where the intrinsic electronic reconstruction has to be disentangled from extrinsic growth related effects.

The research leading to these results has received funding from European Union Seventh Framework Programme (FP7/2007-2013) under grant agreement N. 264098 - MAMA, and from the Italian Ministry of Education, University and Research (MIUR) under Grant Agreement PRIN 2008 - 2DEG FOXI.

**Tables**

Table I: Variation of the LGO/STO heterostructure sheet resistance $R_s^*$ at room temperature, deposition rate, $\delta$, and plume front impact velocity, $u_s$, with the target-to-substrate distance, $d_{TS}$.

| $d_{TS}$ (mm) | $R_s^*$ ($\Omega/\square$) | $\delta$ (Å/shot) | $u_s$ (m/s) |
|---|---|---|---|
| 30 | $1.0\ 10^4$ | 0.26 | $\approx 1.6\times 10^3$ |
| 33 | $1.0\ 10^6$ | 0.21 | $\approx 1.3\times 10^3$ |
| 35 | $6.2\ 10^6$ | 0.18 | $\approx 1.1\times 10^3$ |
| 38 | $\geq 10^9$ | 0.11 | $\approx 0.8\times 10^2$ |

**FIGURE CAPTIONS**

Figure 1: (Color online) (a) Temperature dependence of the sheet resistance, $R_s$, of LGO/STO interfaces grown at oxygen pressure of $10^{-1}$ mbar for a target-substrate distance $d_{TS}$ of 30, 33 and 35 mm, respectively. (b) RHEED intensity monitoring of LGO on STO at $p=10^{-1}$ mbar. STO RHEED pattern before the deposition and final RHEED pattern after growth of 12 uc of LGO are also shown on the top.(c) AFM height image of 12 uc LGO at $p=10^{-1}$ mbar. The cross section corresponding to the straight line in the image is reported on the top.

FIG. 2. (Color online) (a) 2D single shot images of the LGO plume emission at three different oxygen pressure: $10^{-3}$ mbar (left column), $10^{-1}$ mbar (central column), and $10^0$ mbar (right column), for two different delays $\tau$ after the laser pulse. The plume propagation direction is along the $z$-axis, and $z=0$ marks the position of the target surface, while the $x$-axis is parallel to the target surface. To facilitate the comparison, the image intensity is multiplied by an appropriate factor shown in parenthesis in each panel.

FIG. 3. (Color online) (a) Plume front expansion dynamics in dimensionless variables ($p=10^{-3}$ mbar – squares; $p=10^{-2}$ mbar – circles; $p=10^{-1}$ mbar – diamond). The solid line is a fit according to the model described in the text, while the dash-dot line shows the free-plume propagation. (b) Dimensionless coordinates ($\tau,\xi$) corresponding to the deposition conditions used for the fabrication of the LGO/STO heterostructures at $p=10^{-1}$ mbar: star – $d_{TS}$=30 mm; hexagon – $d_{TS}$=33 mm; pentagon – $d_{TS}$=35 mm; triangle – $d_{TS}$=38 mm. The data are shown in a log-log plot. The black line is the fitting curve of panel (a) while the red curve shows a SW-like propagation dynamics ($\xi \propto \tau^{2/5}$).

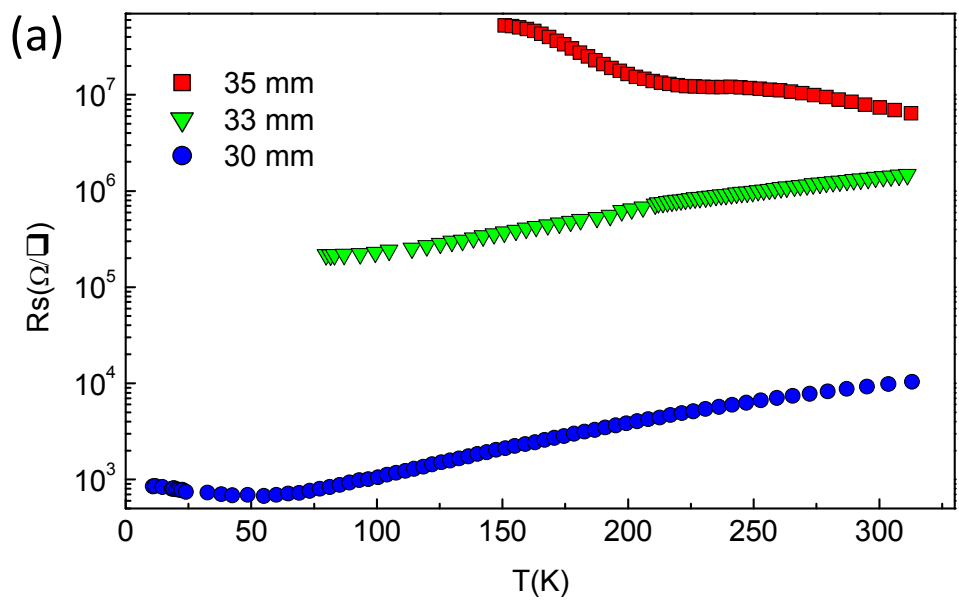
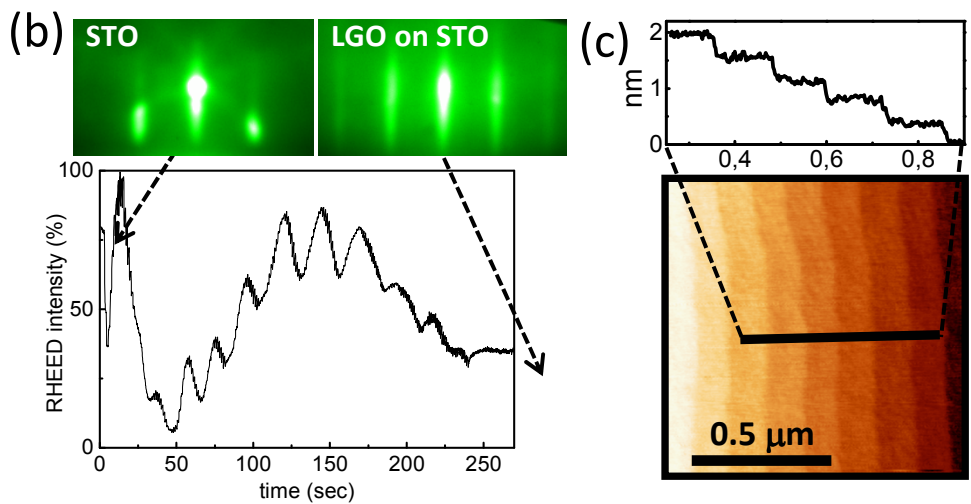

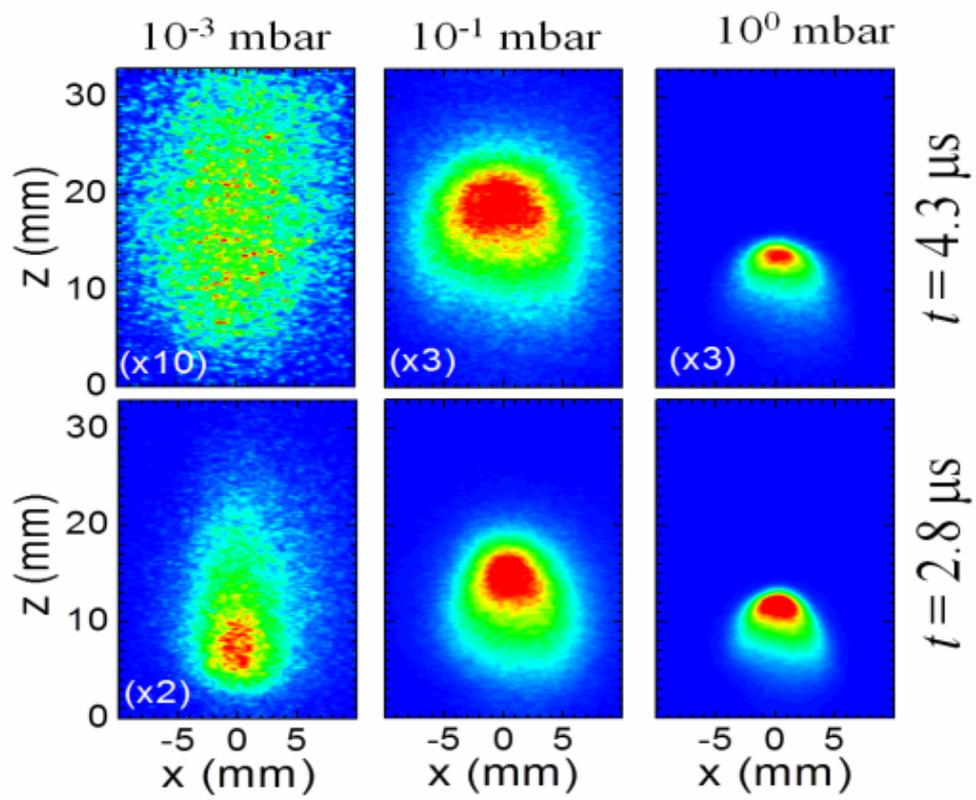

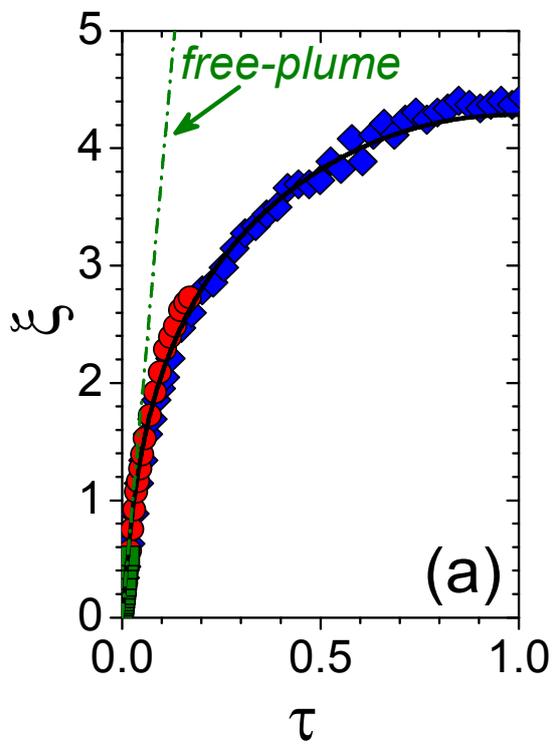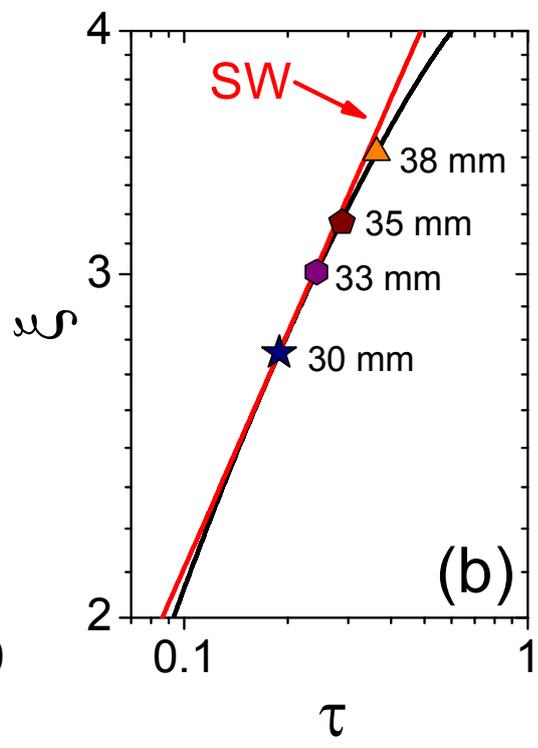